\documentclass[preprint2 ]{aastex62}
\submitjournal{ApJ}

\shorttitle{Sample article}
\shortauthors{Fei et al.}

\begin{document}

\title{The harmonic component of the millihertz quasi-periodic oscillations in 4U 1636$-$53}

\correspondingauthor{}
\email{lvming@xtu.edu.cn}

\author{Zhenyan Fei}
\affiliation{Department of Physics, Xiangtan University, Xiangtan, Hunan 411105, China}
\affiliation{Key Laboratory of Stars and Interstellar Medium, Xiangtan University, Xiangtan, Hunan 411105, China}

\author{Ming Lyu}
\affiliation{Department of Physics, Xiangtan University, Xiangtan, Hunan 411105, China}
\affiliation{Key Laboratory of Stars and Interstellar Medium, Xiangtan University, Xiangtan, Hunan 411105, China}

\author{Mariano M\'endez}
\affiliation{Kapteyn Astronomical Institute, University of Groningen, PO BOX 800, NL-9700 AV Groningen, the Netherlands}

\author{D. Altamirano}
\affiliation{Physics \& Astronomy, University of Southampton, Southampton, Hampshire SO17 1BJ, UK}

\author{Guobao Zhang}
\affiliation{Yunnan Observatories, Chinese Academy of Sciences (CAS), Kunming 650216, P.R. China}
\affiliation{Key Laboratory for the Structure and Evolution of Celestial Objects, CAS, Kunming 650216, P.R. China}

\author{G. C. Mancuso}
\affiliation{Instituto Argentino de Radioastronom\'{\i}a (CCT-La Plata, CONICET; CICPBA), C.C. No. 5, 1894 Villa Elisa, Argentina}
\affiliation{Facultad de Ciencias Astron\'omicas y Geof\'{\i}sicas, Universidad Nacional de La Plata, Paseo del Bosque s/n, 1900 La Plata, Argentina}

\author{Fu-Yuan Xiang}
\affiliation{Department of Physics, Xiangtan University, Xiangtan, Hunan 411105, China}
\affiliation{Key Laboratory of Stars and Interstellar Medium, Xiangtan University, Xiangtan, Hunan 411105, China}

\author{X.J. Yang}
\affiliation{Department of Physics, Xiangtan University, Xiangtan, Hunan 411105, China}
\affiliation{Key Laboratory of Stars and Interstellar Medium, Xiangtan University, Xiangtan, Hunan 411105, China}




\begin{abstract}

We studied the harmonics of the millihertz quasi-periodic oscillations (mHz QPOs) in the neutron-star low-mass X-ray binary 4U 1636--53 using the Rossi X-ray Timing Explorer observations. We detected the harmonics of the mHz QPOs in 73 data intervals, with most of them in the transitional spectra state. We found that the ratio between the rms amplitude of the harmonic and that of the fundamental remains constant in a wide range of the fundamental frequency. More importantly, we studied, for the first time, the rms amplitude of the harmonics vs. energy in 4U 1636--53 in the 2-5 keV range. We found that the rms amplitude of both the harmonic and the fundamental shows a decreasing trend as the energy increases, which is different from the behaviors reported in QPOs in certain black hole systems. Furthermore, our results suggest that not all observations with mHz QPOs have the harmonic component, although the reason behind this is still unclear.

\end{abstract}



\section{Introduction}

Milli-hertz quasi-periodic oscillations (mHz QPOs) were discovered in the neutron-star low-mass X-ray binaries 4U 1656--53, 4U 1608--52 and Aql X--1 \citep{Revnivtsev2001}. The mHz QPOs in these systems show some unique properties \citep{Revnivtsev2001,Altamirano2008b,Lyu2015}: (1) the oscillation frequencies are always  below $\sim$ 14 mHz; (2) the QPOs are detected only when the source is within a certain luminosity range, $L_{\rm 2-20\:keV} \simeq (5-11) \times 10^{36}$ ergs s$^{-1}$; (3) the QPOs are significantly detected only in the soft X-ray band (\begin{math}<\end{math} 5 keV); (4) the QPOs become undetectable when a type I X-ray burst happens.

\citet{Revnivtsev2001} speculated that the mHz QPOs could be related to a special mode of nuclear burning on the neutron-star surface. \citet{Heger2007} found that marginally stable nuclear burning of helium on the neutron-star surface could produce the observed mHz QPOs. However, in the simulations of \citet{Heger2007}, the mHz QPOs are present only when the mass accretion rate is close to the Eddington rate, an order of magnitude higher than the average global accretion rate deduced from observations. To solve this discrepancy, \citet{Heger2007} proposed that the local accretion rate in the burning layer where mHz QPOs happen could be higher than the global rate. \citet{keek2009} found that turbulent chemical mixing of the fuel, together with a higher heat flux from the crust, is able to explain the observed accretion rate at which mHz QPOs are seen. Furthermore, \citet{keek2014} found that the mHz QPOs could not be triggered at the observed accretion rates by changing only the chemical composition and the nuclear reaction rate.

The scenario in which the mHz QPOs originate from nuclear burning on the neutron-star surface was supported by other observations. \citet{Yuvan2002} found that in 4U 1608--52 the frequency of a kilohertz (kHz) QPO was anti-correlated with the 2-5 keV count rate when a mHz QPO was present, supporting the nuclear burning origin of the mHz QPOs; in their interpretation, the accretion disk is pushed outward from the neutron-star surface by radiation stress in each mHz QPO cycle, leading to the frequency variation of the kHz QPO. \citet{Altamirano2008b} found that the frequency of the mHz QPOs in 4U 1636--53 systematically decreased before a type I X-ray burst, and then the QPOs disappeared when a burst occurred. Recently, the same behavior was detected in another mHz QPO source, LMXB EXO 0748--676 \citep{Mancuso2019}. The frequency drift behavior of the mHz QPOs indicates that the mHz QPOs are closely related to nuclear burning on the neutron-star surface. Furthermore, a smooth evolution between the mHz QPOs and the X-ray bursts was reported by \citet{Linares2012} in IGR J17480--2446: as the accretion rate increased, bursts gradually evolved into a mHz QPO, and vice versa. 

Recently, other properties of the mHz QPOs have been investigated. \citet{Lyu2015} found that the frequency of the mHz QPOs was not significantly correlated with the temperature of the neutron-star surface in 4U 1636--53, different from the model predictions. They also found that the mHz QPOs are likely connected to He-rich X-ray bursts. A study of the shape of X-ray bursts connected to mHz QPOs indicates that these bursts have positive convexity and short rising time, suggesting that the mHz QPOs originate on the equatorial region of the neutron-star surface \citep{Lyu2016}. \citet{Lyu2019} showed that the fractional rms amplitude of the mHz QPOs is anti-correlated with the count rate below 5 keV, and the absolute RMS amplitude of the mHz QPOs is insensitive to the parameter $S_{a}$ that measures the position of the source in the color-color diagram. Using XMM-Newton and Neutron Star Interior Composition Explorer observations, \citet{Lyu2020} found that the rms amplitude of the mHz QPOs in 4U 1636--53 first increases from $\sim$ 0.2 keV to $\sim$ 3 keV, and then decreases above $\sim$ 3 keV. Besides, \citet{Stiele2016} studied phase-resolved energy spectra of the mHz QPOs in 4U 1636--53, and concluded that the QPOs were not due to the modulations of the neutron-star surface temperature. On the contrary, \citet{Strohmayer2018} recently found that in GS 1826--238 the mHz oscillations are consistent with variation of the blackbody temperature of the neutron-star surface, although other interpretations were possible.

In the past, the main focus has been to understand the fundamental of the mHz QPOs, while the harmonic component has received less attention. There is limited description of the harmonic in the current theories of the mHz QPOs. Simulations in the work of \citet{Heger2007} indicate that the profile of the oscillations is asymmetric, with the decay lasting twice as long as the rise. This feature is consistent with the findings in \citet{Revnivtsev2001} that the profile of the mHz QPOs were asymmetric. \citet{Heger2007} proposed that the asymmetric profile should be due to the existence of significant harmonic components. Notwithstanding, \citet{Strohmayer2018} recently found the shape of the mHz oscillation in GS 1826--238 observed with NICER is quite symmetric.

In this work, for the first time, we focus on the properties of the harmonics of the mHz QPOs in the LMXB 4U 1636--53 using observations with the Rossi X-ray Timing Explorer (RXTE). The paper is organized as follow: In Section 2 we describe the observations and data analysis; in Section 3 we present our main results. Finally, we briefly discuss our results in Section 4.

\section{Observations and data reduction}

We analyzed the RXTE observations of 4U 1636--53 presented in \citet{Lyu2019} using the Proportional Counter Array \citep[PCA;][]{Jahoda2006} onboard RXTE. We extracted light curves of 1s resolution in the $\sim$2-4.5 keV energy range from all available proportional counter units (PCUs). We divided each light curve into a series of independent 700 s intervals, following exactly the same steps as the ones in the work of \citet{Lyu2019}. We first selected those intervals in \citet{Lyu2019} where either the mHz QPO is 3-$\sigma$ significant in at least two consecutive intervals, or there is a harmonic at twice the QPO frequency if there is only one interval available. We then searched the harmonic components in these intervals using the Lomb-Scargle periodograms \citep{Lomb1976, Scargle1982}, and applied a 3-$\sigma$ confidence level for the detection. We set the number of trials as the number of independent frequencies in the Lomb-Scargle periodogram in each interval to calculate the 3-$\sigma$ detection level for the QPOs, and to one to calculate the 3-$\sigma$ level for the harmonics. The reason for the latter is that the frequency of the harmonic is known a priori once the frequency of the fundamental is known. In Figure 1, we show an example of the detection of the fundamental and the second harmonic of the mHz QPO in the Lomb-Scargle periodogram.

For each interval with a harmonic, we folded the corresponding light curve using the tool {\tt efold}. The period for the folding is derived from the Lomb-Scargle periodograms: For this we took the frequency at which the power is the maximum in the periodogram as the oscillation frequency of the fundamental of the mHz QPO in that interval \citep{Lyu2019} and folded the light curves at that period. We then fitted each folded light curve with the model {\tt constant+sine1+sine2}. The constant component accounts for the persistent flux in the light curve, and the {\tt sine1} and {\tt sine2} describe the structure of the fundamental and the harmonic. Considering that each light curve has already been folded at the QPO period, we fixed the period of the {\tt sine1} and {\tt sine2} component in the model at 1 and 0.5, respectively. We estimated the background with the tool {\tt pcabackest} and found that the background count rate divided by the total rate is less than $\sim$ 3\%. We therefore did not include the background into the following calculation since it is much smaller than the error of the fractional rms amplitude. Finally, we calculated the fractional rms amplitude of the fundamental and the harmonic with the formula $rms_{f}=A1 /[\sqrt{2} \times C]$ and $rms_{h}=A2 /[\sqrt{2} \times C]$, where A1 and A2 are the amplitude of the {\tt sine1} and the {\tt sine2} functions, respectively, and C is the value of the constant component. We also calculated the absolute RMS amplitude of the fundamental and the harmonic, $RMS_{f}=rms_{f} \times R$ and $RMS_{h}=rms_{h} \times R$, where R is the average count rate measured by the PCU2 detector.

To explore the rms spectra of the fundamental and the harmonic, we further extracted light curves in four different energy bands, $\sim$ 2.0-2.8 keV, $\sim$ 2.8-3.7 keV, $\sim$ 3.7-4.5 keV, $\sim$ 4.5-5.4 keV, for each 700s interval. We then applied the same folding and the fitting procedure as the one described above, and calculated the fractional rms amplitude of the fundamental and harmonic at different energies.

Besides, we constructed a color-color diagram to trace the spectral state of the source during the observations where the harmonics are present. For this we used the 16 s time-resolution Standard-2 data available from the RXTE/PCA. The soft color was calculated as the count rate in the 3.5-6.0 keV band divided by that in the 2.0-3.5 keV band, and the hard color was defined as the ratio between the rate in the 9.7-16.0 keV and the one in the 6.0-9.7 keV. We normalized the colors to those of the Crab Nebula in observations taken close in time to those of 4U 1636--53 \citep[see][for more details]{Altamirano2008a, Zhang2009}. To locate the position of the observation in the color-color diagram, we further calculated the parameter $S_{a}$ as in \citet{Zhang2011}: the value of the $S_{a}$ is normalized to the distance between $S_{a}$ = 1 at the top right-hand vertex and $S_{a}$ = 2 at the bottom left-hand vertex in the color-color diagram.

\section{Results}

In total, we detected the harmonic component of the mHz QPOs in 73 individual intervals. In Figure \ref{ccd} we show the distribution of the observations with harmonics in the color-color diagram, together with the distribution of the observations with mHz QPO reported in \citet{Lyu2019}. We found that almost all the observations with harmonics are located in the transitional spectral state between the soft and the hard state, with only one exception in the very soft state. These observations cluster in a small region in the plot where the soft and the hard color are around 1.0 and 0.6-0.8, respectively. The details of these 73 intervals with the harmonics are shown in Table 1.

\begin{figure*}
\centering
\includegraphics[width=0.5\textwidth]{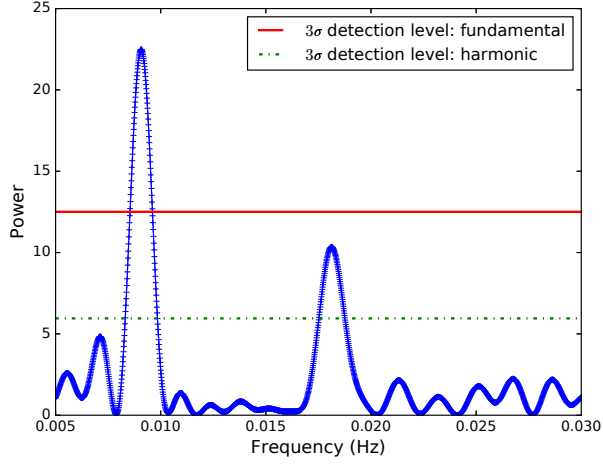}
\caption{Lomb-Scargle periodogram of the data interval (ObsID: 40028-01-19-00) with the mHz QPO and its harmonic component. We oversampled the frequency by a factor of 100, and marked the 3-$\sigma$ significance detection level for the fundamental (red horizontal line) and the harmonic (green horizontal dotted line), respectively (See text for the explanation of the difference between the red and the green dotted lines).} 
\label{LS}
\end{figure*}

\begin{figure*}
\centering
\includegraphics[width=0.5\textwidth]{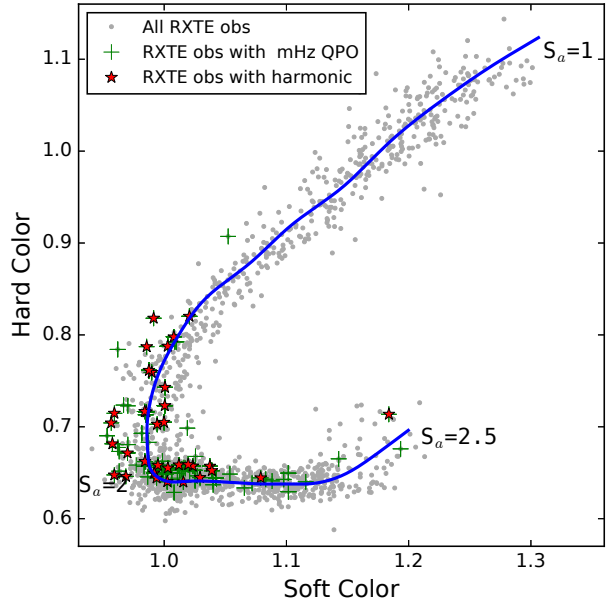}
\caption{Color-color diagram of 4U 1636--53 with RXTE. We used gray points to represent the averaged Crab-normalized colors of RXTE observations \citep[see][for more details]{Zhang2011}. Each point corresponds to one RXTE observation. The red stars in the figure show the position of the observations where mHz QPOs show harmonics. The green crosses correspond to the observations with the mHz QPOs reported in \citet{Lyu2019}. We used the length of the blue curve $S_{a}$ to parametrize the position of the source in this diagram.}
\label{ccd}
\end{figure*}

\begin{figure*}
\centering
\includegraphics[width=0.6\textwidth]{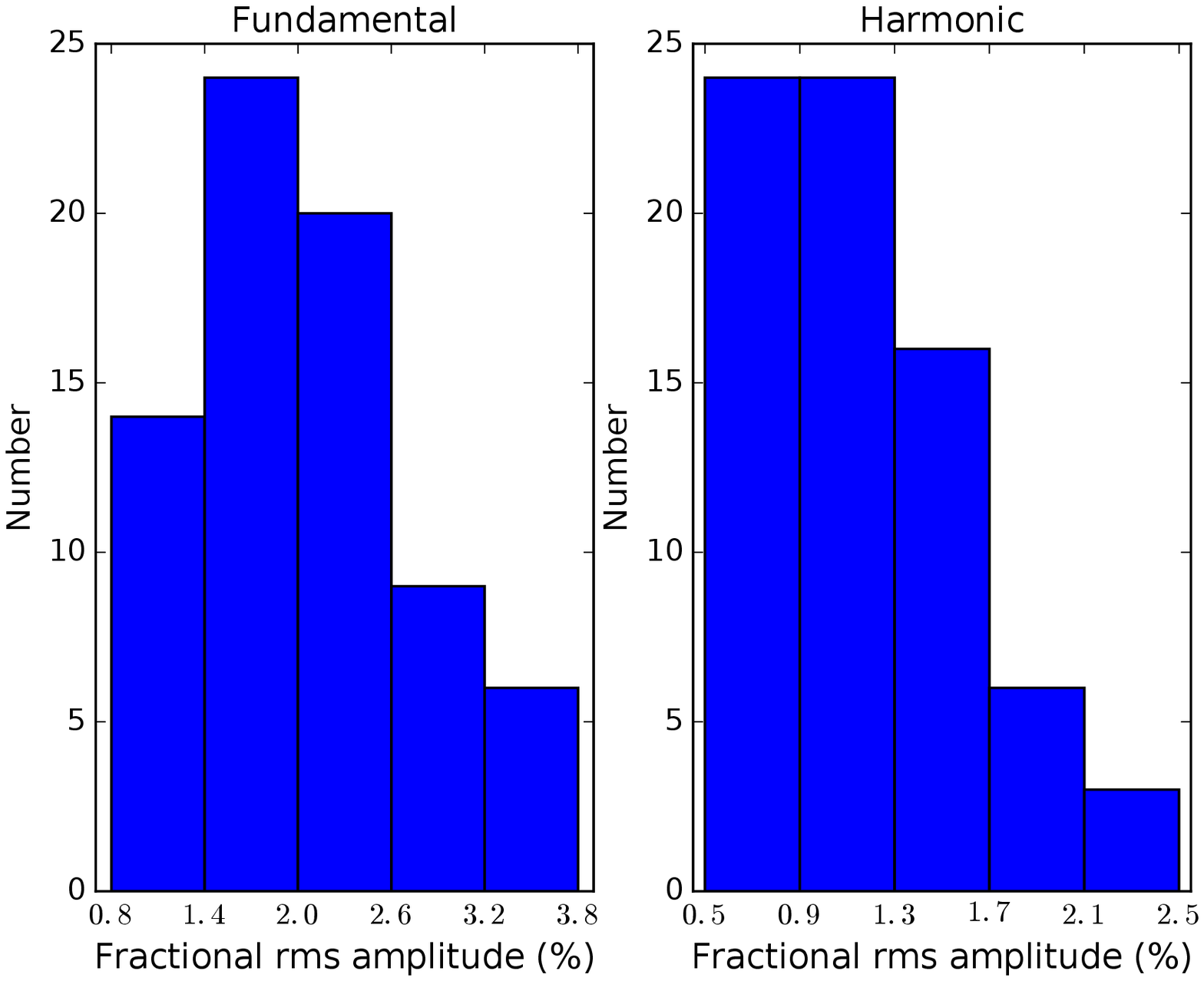}
\caption{Distribution of the fractional rms amplitude of the fundamental (left) and the harmonic (right) of the mHz QPO in 4U 1636--53. The rms amplitudes are measured in the 2-4.5 keV energy range.}
\label{hist_Frac_rms}
\end{figure*}

\begin{figure*}
\centering
\includegraphics[width=0.61\textwidth]{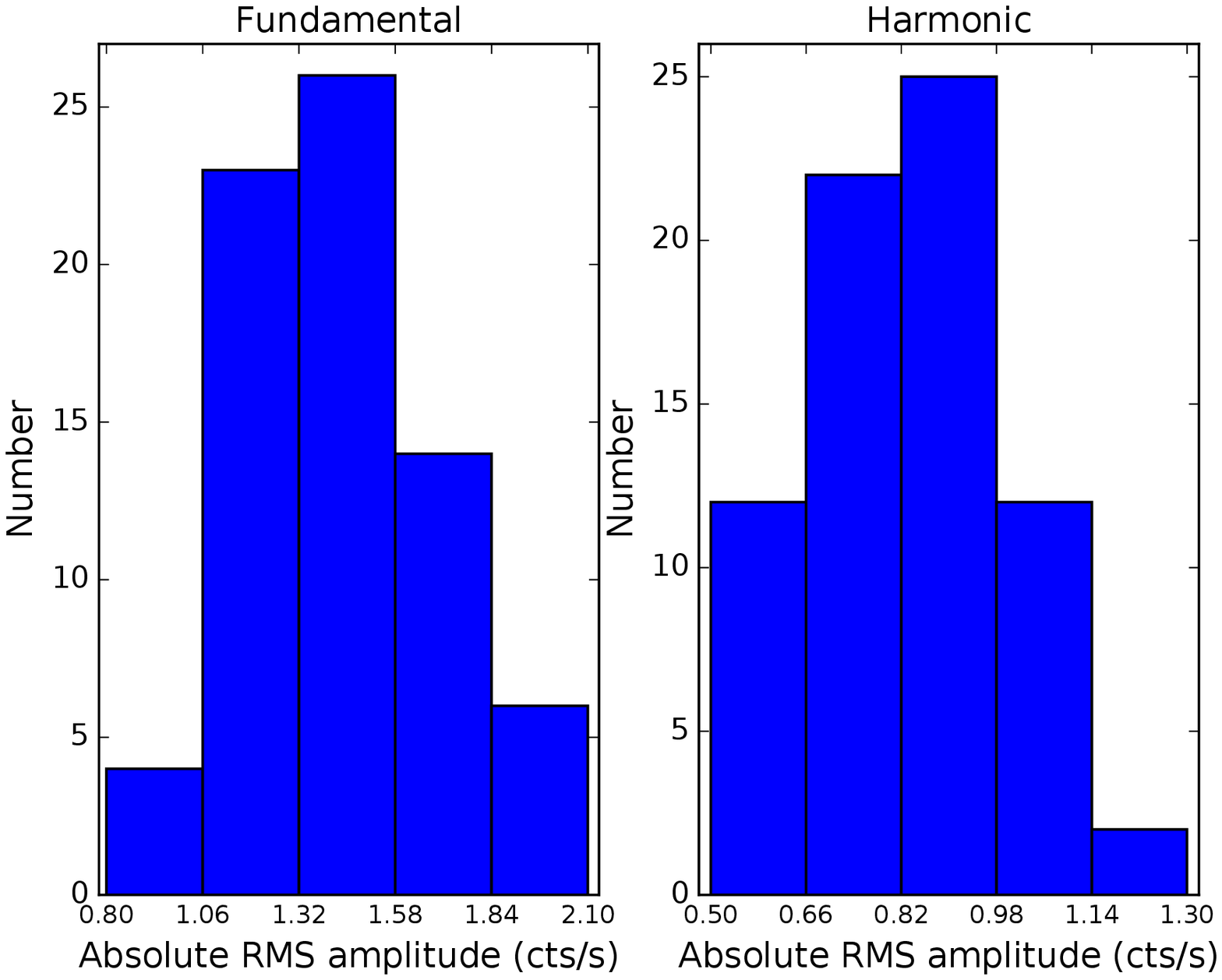}
\caption{Distribution of the absolute RMS amplitude of the fundamental (left) and the harmonic (right) of the mHz QPO in 4U 1636--53. The RMS amplitudes are measured in the 2-4.5 keV energy range.}
\label{hist_abs_rms}
\end{figure*}

\begin{figure*}
\centering
\includegraphics[width=0.5\textwidth]{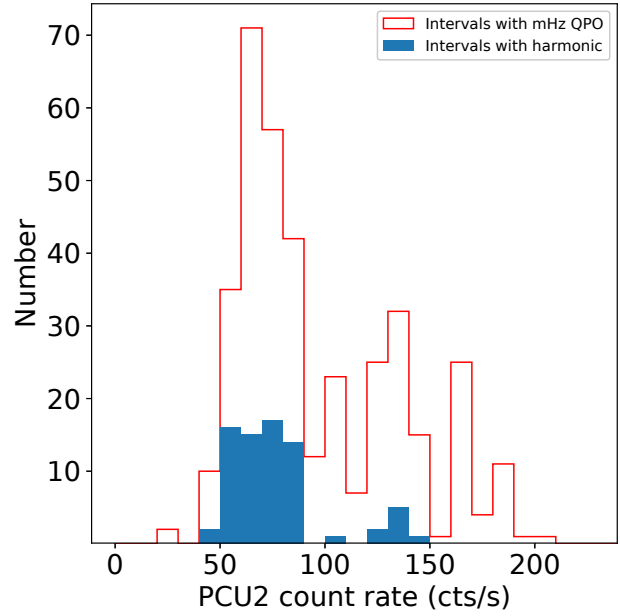}
\caption{Distribution of the PCU2 count rate of the intervals with the mHz QPOs and the harmonics in 4U 1636--53. The count rate is measured below $\sim$ 5 keV.}
\label{hist_rate}
\end{figure*}

\begin{figure*}
\centering
\includegraphics[width=0.45\textwidth]{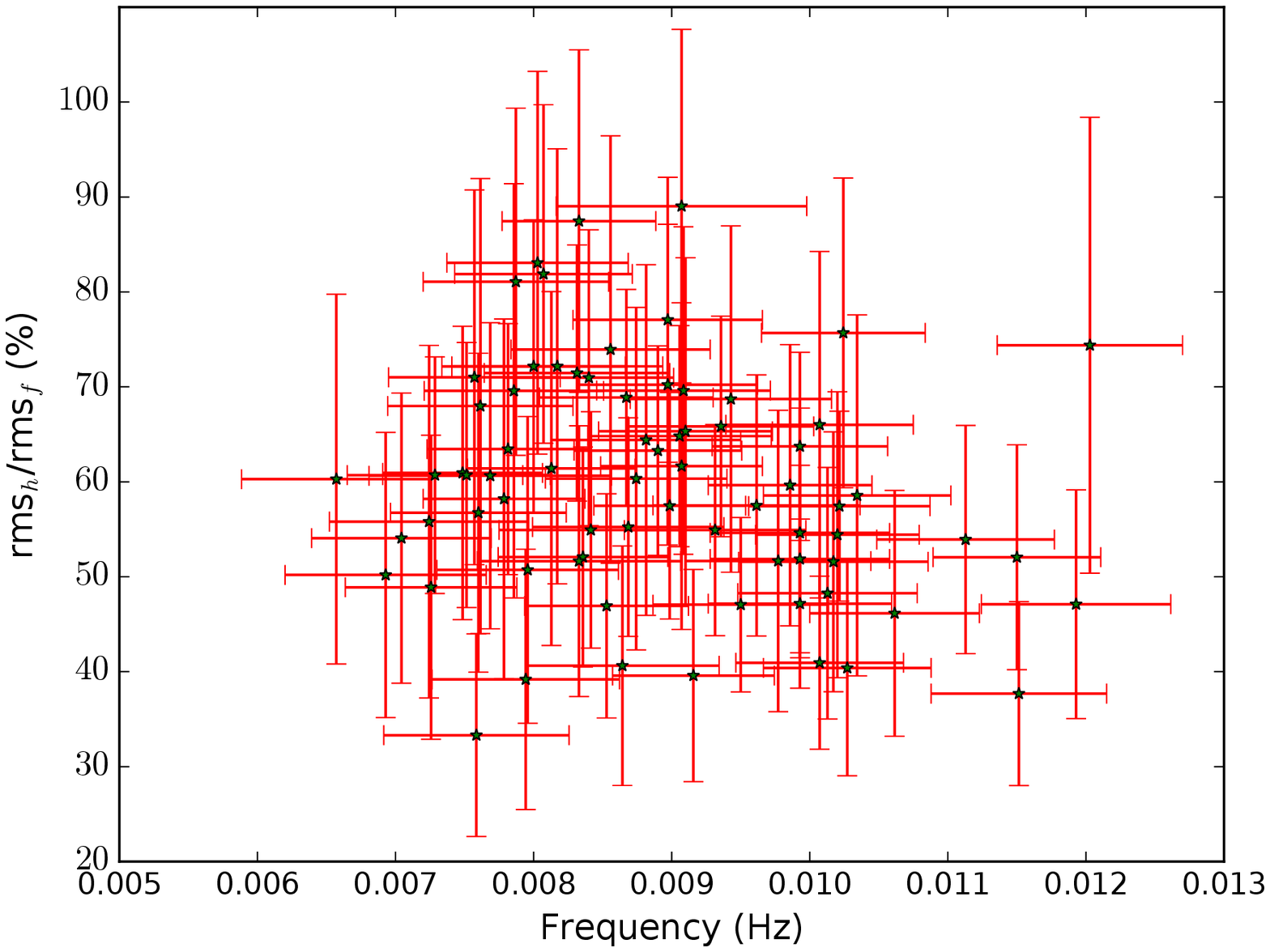}
\includegraphics[width=0.45\textwidth]{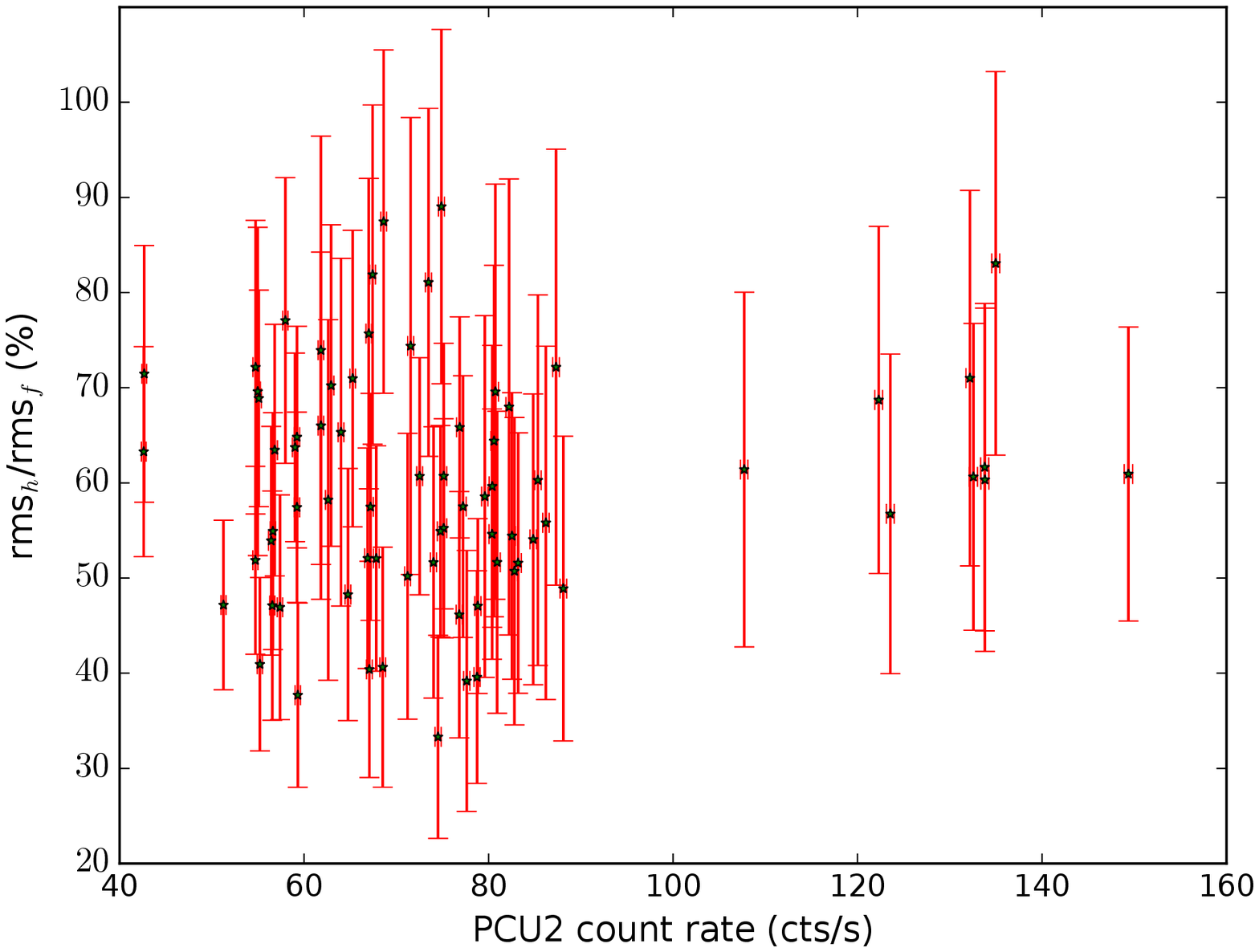}
\caption{The ratio between the fractional rms amplitude of the harmonic and that of the fundamental vs. the frequency of the fundamental (left panel) and the PCU2 count rate below 5 keV (right panel) in each interval in 4U 1636--53.}
\label{ratio}
\end{figure*}

\begin{figure*}
\centering
\includegraphics[width=0.6\textwidth]{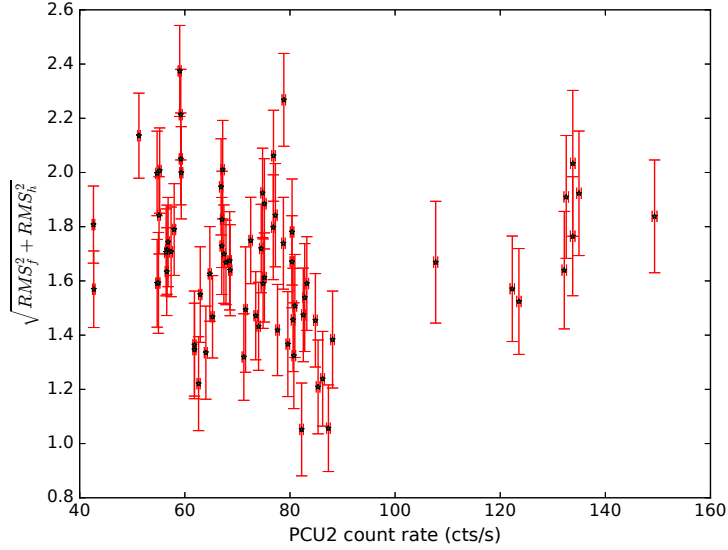}
\caption{The square root of the sum of the square of the absolute RMS amplitude of the fundamental and the harmonic of the mHz QPO in 4U 1636--53 vs. the PCU2 count rate below 5 keV in each interval.}
\label{t}
\end{figure*}

\begin{table*}
\centering
\caption{RXTE observations of the harmonics of the mHz QPOs in 4U 1636--53. For clarity, we marked the different datasets in the same observation with different labels, for instance, ’D1’ and ’D2’ in observation 60032-01-19-000.}

\resizebox{\textwidth}{!}{
\scalebox{0.90}{
\begin{tabular}{lccclccc}
\hline
Observation ID       & Dataset Start Time            & Intervals (s) 				&QPO Frequency (mHz) &Observation ID		 & Dataset Start Time 		& Intervals (s)		&QPO Frequency (mHz) \\
\hline
30053-02-01-000	&1998-02-25 04:22:24	&700-1400	&9.43$\pm$0.73	&60032-05-06-00	&2002-01-14 07:05:36	&9100-9800	&10.27$\pm$0.61	\\
30053-02-01-001	&1998-02-25 05:58:24	&13300-14000	&7.69$\pm$0.88	&60032-05-06-00	&2002-01-14 07:05:36	&10500-11200	&10.24$\pm$0.59\\
30053-02-01-001	&1998-02-25 05:58:24	&17500-18200	&8.03$\pm$0.66	&60032-05-06-00	&2002-01-14 07:05:36	&12600-13300	&8.99$\pm$0.55\\
30053-02-02-02	&1998-08-19 08:15:06	&0-700	&7.60$\pm$0.64	&	60032-05-06-00	&2002-01-14 07:05:36	&14000-14700	&8.64$\pm$0.70\\
40028-01-04-00	&1999-04-28 23:57:09	&1400-2100	&8.74$\pm$0.66	&60032-05-06-00	&2002-01-14 07:05:36	&15400-16100	&8.36$\pm$0.61\\
40028-01-15-00	&2000-06-15 03:49:32	&700-1400	&7.49$\pm$0.58	&60032-05-06-00	&2002-01-14 07:05:36	&16100-16800	&8.07$\pm$0.64\\
40028-01-19-00	&2000-08-12 20:22:18	&8400-9100	&9.07$\pm$0.59	&60032-05-07-01 (G1) 	&2002-01-15 05:21:20	&700-1400	&11.93$\pm$0.69\\
60032-01-01-01	&2001-06-14 22:33:40	&700-1400	&7.86$\pm$0.65	&60032-05-07-01 (G1) 	&2002-01-15 05:21:20	&2100-2800	&11.13$\pm$0.64\\
60032-01-01-01	&2001-06-14 22:33:40	&1400-2100	&7.96$\pm$0.66	&60032-05-07-01 (G1) 	&2002-01-15 05:21:20	&2800-3500	&10.07$\pm$0.61\\
60032-01-08-00	&2001-08-28 14:12:32	&1400-2100	&7.57$\pm$0.62	&60032-05-07-01 (G1) 	&2002-01-15 05:21:20	&3500-4200	&9.93$\pm$0.65\\
60032-01-11-03	&2001-09-17 06:08:32	&2100-2800	&11.50$\pm$0.61	&60032-05-07-01 (G2) 	&2002-01-15 06:49:36	&0-700	&8.67$\pm$0.63\\
60032-01-19-000 (D1) 	&2002-01-08 06:57:51	&0-700	&7.26$\pm$0.62	        &60032-05-09-00	&2002-01-15 22:37:11	&1400-2100	&8.40$\pm$0.61\\
60032-01-19-000 (D2)	&2002-01-08 08:34:24	&0-700	&7.04$\pm$0.65	        &60032-05-23-000	&2003-01-07 13:06:40	&8400-9100	&10.61$\pm$0.61\\
60032-01-19-000 (D2)	&2002-01-08 08:34:24	&700-1400	&7.24$\pm$0.72	&60032-05-23-000	&2003-01-07 13:06:40	&9100-9800	&9.61$\pm$0.75\\
60032-01-19-000 (D2)	&2002-01-08 08:34:24	&1400-2100	&6.57$\pm$0.69	&60032-05-23-000	&2003-01-07 13:06:40	&10500-11200	&10.34$\pm$0.68\\
60032-01-20-000 (E1)	&2002-01-08 21:11:50	&0-700	&9.77$\pm$0.67	        &60032-05-23-000	&2003-01-07 13:06:40	&11200-11900	&9.86$\pm$0.59\\
60032-01-20-000 (E1)	&2002-01-08 21:11:50	&1400-2100	&8.81$\pm$0.69	        &60032-05-24-00 (H1)  	&2003-01-07 03:47:28	&0-700	&9.16$\pm$0.59\\
60032-01-20-000 (E2)	&2002-01-08 22:29:20	&2100-2800	&7.87$\pm$0.67	&60032-05-24-00 (H1) 	&2003-01-07 03:47:28	&700-1400	&9.50$\pm$0.64\\
60032-01-21-00 (F1) 	&2002-01-10 06:30:24	&2100-2800	&10.17$\pm$0.69	&60032-05-24-00 (H1) 	&2003-01-07 03:47:28	&1400-2100	&9.36$\pm$0.67\\
60032-01-21-00 (F1) 	&2002-01-10 06:30:24	&2800-3500	&10.20$\pm$0.59	&60032-05-24-00 (H1) 	&2003-01-07 03:47:28	&2100-2800	&9.07$\pm$0.91\\
60032-01-21-00 (F1) 	&2002-01-10 06:30:24	&3500-4200	&9.93$\pm$0.65	&60032-05-24-00 (H1) 	&2003-01-07 03:47:28	&2800-3500	&8.69$\pm$0.69\\
60032-01-21-00 (F2) 	&2002-01-10 08:06:24	&1400-2100	&7.94$\pm$0.68	&60032-05-24-00 (H2)      &2003-01-07 05:25:20	&3500-4200	&8.33$\pm$0.74\\
60032-01-22-00	&2002-01-10 20:23:28	&700-1400	&7.59$\pm$0.67	&60032-05-24-01 (I1) 	&2003-01-06 04:07:01	&2100-2800	&9.31$\pm$0.66\\
60032-01-22-00	&2002-01-10 20:23:28	&1400-2100	&7.51$\pm$0.61	&60032-05-24-01 (I2) 	&2003-01-06 05:44:32	&1400-2100	&7.29$\pm$0.64\\
60032-01-24-00	&2003-01-03 01:48:32	&0-700	&8.13$\pm$0.61	        &60032-05-24-01 (I2) 	&2003-01-06 05:44:32	&3500-4200	&6.93$\pm$0.73\\	
60032-05-01-00	&2002-01-11 23:29:36	&700-1400	&9.93$\pm$0.66	&70036-01-01-00	&2002-06-10 01:18:24	&0-700	&7.61$\pm$0.67\\
60032-05-02-00	&2002-01-12 09:56:32	&2800-3500	&10.13$\pm$0.65	&91024-01-11-10	&2005-10-07 06:45:36	&0-700	&9.10$\pm$0.63\\
60032-05-02-00	&2002-01-12 09:56:32	&6300-7000	&8.97$\pm$0.65	&91024-01-36-00	&2005-05-14 01:06:40	&0-700	&8.90$\pm$0.61\\
60032-05-02-00	&2002-01-12 09:56:32	&10500-11200	&7.81$\pm$0.59	&91024-01-47-00	&2005-06-05 11:27:10	&0-700	&8.17$\pm$0.76\\
60032-05-04-00	&2002-01-13 07:51:45	&8400-9100	&11.51$\pm$0.64	&91024-01-77-10	&2006-02-16 18:24:32	&700-1400	&8.31$\pm$0.67\\
60032-05-04-00	&2002-01-13 07:51:45	&9800-10500	&10.21$\pm$0.66	&92023-01-07-20	&2007-03-31 14:52:32	&0-700	&7.79$\pm$0.59\\
60032-05-04-00	&2002-01-13 07:51:45	&10500-11200	&9.93$\pm$0.64	&92023-02-17-00	&2007-03-01 01:14:24	&0-700	&8.33$\pm$0.56\\
60032-05-04-00	&2002-01-13 07:51:45	&11900-12600	&9.06$\pm$0.66	&93087-01-47-00	&2007-09-29 11:05:36	&0-700	&10.07$\pm$0.68\\
60032-05-04-00	&2002-01-13 07:51:45	&12600-13300	&8.97$\pm$0.69	&93087-01-58-00	&2007-10-21 03:05:36	&700-1400	&8.56$\pm$0.72\\
60032-05-04-00	&2002-01-13 07:51:45	&13300-14000	&8.53$\pm$0.59	&94087-01-10-00	&2009-01-13 07:41:56	&700-1400	&9.09$\pm$0.63\\
60032-05-04-00	&2002-01-13 07:51:45	&14000-14700	&8.41$\pm$0.66	&94087-01-98-00	&2009-07-08 11:30:58	&0-700	&12.03$\pm$0.67\\
60032-05-04-00	&2002-01-13 07:51:45	&16100-16800	&8.00$\pm$0.66	\\
			
\hline
\end{tabular}}}
\end{table*}

In Figure \ref{hist_Frac_rms} we show the distribution of the fractional rms amplitude of the fundamental and the harmonic components. The rms amplitude of the fundamental distributes in a wide range, ranging from 0.8\% to 3.8\%. The rms amplitude of the harmonic is in the range of 0.5\% to 2.5\%, with most of them in the 0.5\%-1.7\% range. In Figure \ref{hist_abs_rms} we show the distribution of the absolute RMS amplitude of the fundamental and the harmonic measured by the PCU2 detector. The absolute RMS amplitude of the fundamentals distributes in the range of 0.80-2.10 cts/s, with 67\% of them clustering in the narrow range between 1.06 cts/s to 1.58 cts/s. The range of the absolute RMS amplitude of the harmonic is a bit narrower, distributing in the range of 0.5-1.3 cts/s.

In Figure \ref{hist_rate} we show the distribution of the PCU2 count rate below $\sim$ 5 keV for the intervals with the mHz QPOs in \citet{Lyu2019} and the ones with the harmonics detected in this work. The intervals with the harmonics have a count rate between 40 to 150 cts/s, narrower than the range covered by the intervals with the mHz QPOs. We applied a Kolmogorov-Smirnov test to test whether the two distribution are consistent with being drawn from the same parent distribution. The null hypothesis of the Kolmogorov-Smirnov test is that the two independent samples are drawn from the same parent population. The p-value from the Kolmogorov-Smirnov test here is very low, around $10^{-5}$, suggesting that we can reject the null hypothesis, so the two count rate samples are likely not from the same parent distribution.

In the left panel of Figure \ref{ratio} we show the ratio between the rms amplitude of the harmonic and that of the fundamental in each interval vs. the frequency of the fundamental. The fundamental frequency is in the range of $\sim$6-12 mHz, and we found that there is no correlation between the rms ratio and the fundamental frequency. In the right panel we show the rms amplitude ratio as a function of the PCU2 count rate below $\sim$5 keV. We found no correlation between these two parameters either. The rms amplitude ratio in these 73 intervals are consistent with being constant around a value of 60\% within the errors. The averaged ratio and its standard deviation are 0.6 and 0.1, respectively. In Figure \ref{t} we show the square root of the sum of the square of the absolute RMS amplitudes of the fundamental and harmonic as a function of the PCU2 count rate. We fitted a linear function $y=a \times x+b$ to the data and found that the parameter $a$ is less than 3$\sigma$ different from zero, indicating that these two quantities are not correlated.

In Figure \ref{res1} we plot the rms amplitude of the fundamental and its harmonic as a function of energy in different intervals. We found that the rms amplitude of the fundamental in these intervals shows a decreasing trend as the energy increases from $\sim$ 3 keV up to $\sim$ 5.4 keV. The rms amplitude of the harmonic decreases or remains more or less constant within the errors as the energy increases.

To explore whether the rms amplitude of the fundamental and the harmonic follow the same trend as the energy increases, we further test the rms spectra of the fundamentals and the harmonics in three ways:

(a) We subtracted the rms spectra of the fundamental and that of the harmonic in each interval by their average, and calculated the difference at each energy between the average-subtracted rms amplitude of the fundamental and the harmonic. 

(b) We divided all the rms spectra of the fundamental and that of the harmonic by their average to bring them to the same scale, and then compared the normalized rms spectra in each interval.

(c) We divided the rms spectra of the harmonics by that of the fundamentals, and then fitted the ratio with a function $y=a \times x+b$. If the two rms spectra have the same trend, the ratio should be some constant value, and the slope $a$ should be consistent with zero. 

We found that the absolute value of the difference between the two average-subtracted rms spectra divided by its 1$\sigma$ error at each energy is always less than 3, suggesting that the rms spectra of the fundamentals and the harmonics have the same trend. Furthermore, we found that the normalized rms spectra of the fundamental and that of the harmonic in each interval are consistent with each other within the 3$\sigma$ errors. Finally, the fitted slope $a$ in the 73 intervals in this work are all consistent with zero within 3 sigma errors. 

In Figure \ref{sploe_nor} we show the normalized rms spectra of the fundamental and the harmonic in the 73 samples. The normalized rms amplitude of the fundamental shows a clear decreasing trend as the energy increases, and the rms amplitude of the harmonic shows a similar tendency. We fitted the rms spectra with a linear function and found that the slope of the fundamentals and the harmonics is $-0.30\pm0.04$ (3-$\sigma$ level) and $-0.26\pm0.07$, respectively.

\begin{figure*}
\centering
\includegraphics[height=115mm,width=72mm]{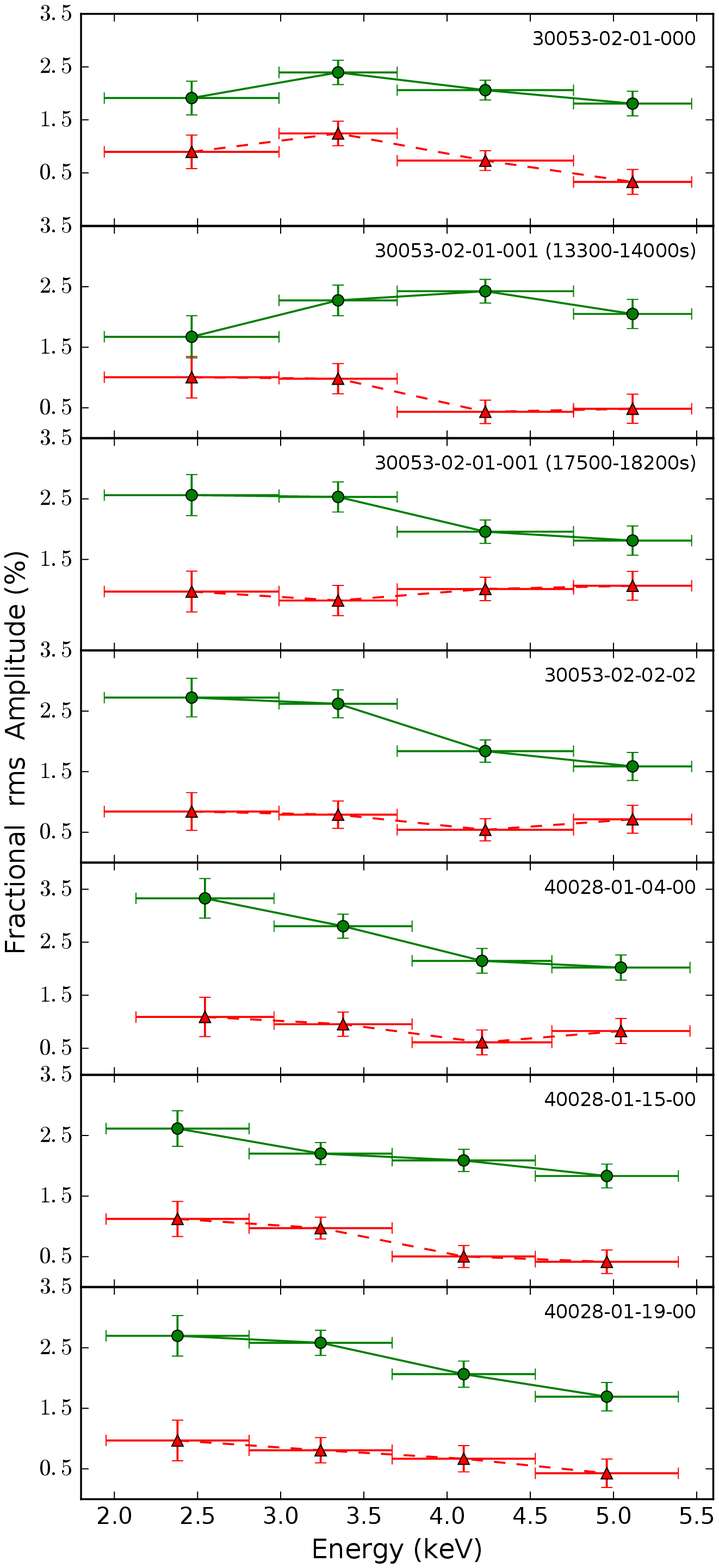}
\includegraphics[height=115mm,width=72mm]{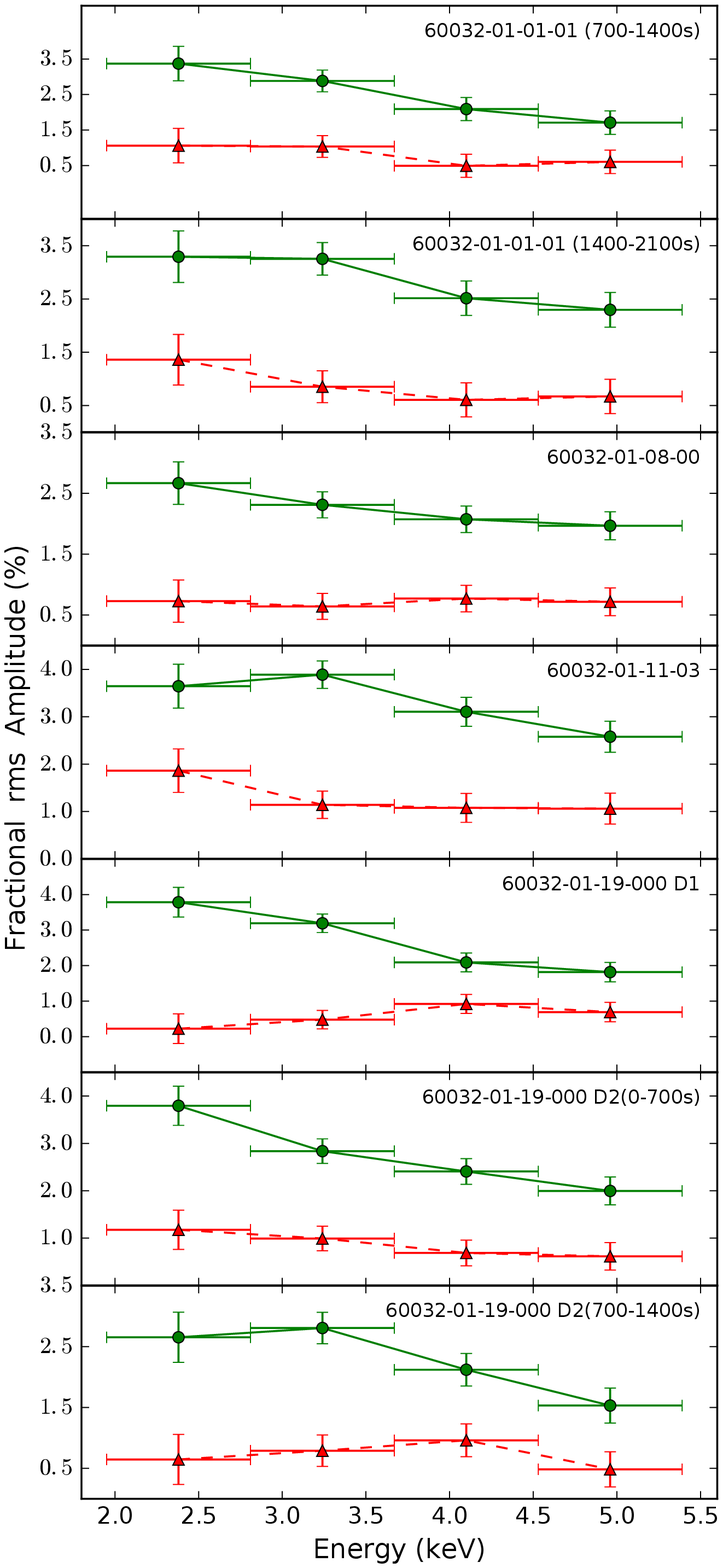}
\includegraphics[height=115mm,width=72mm]{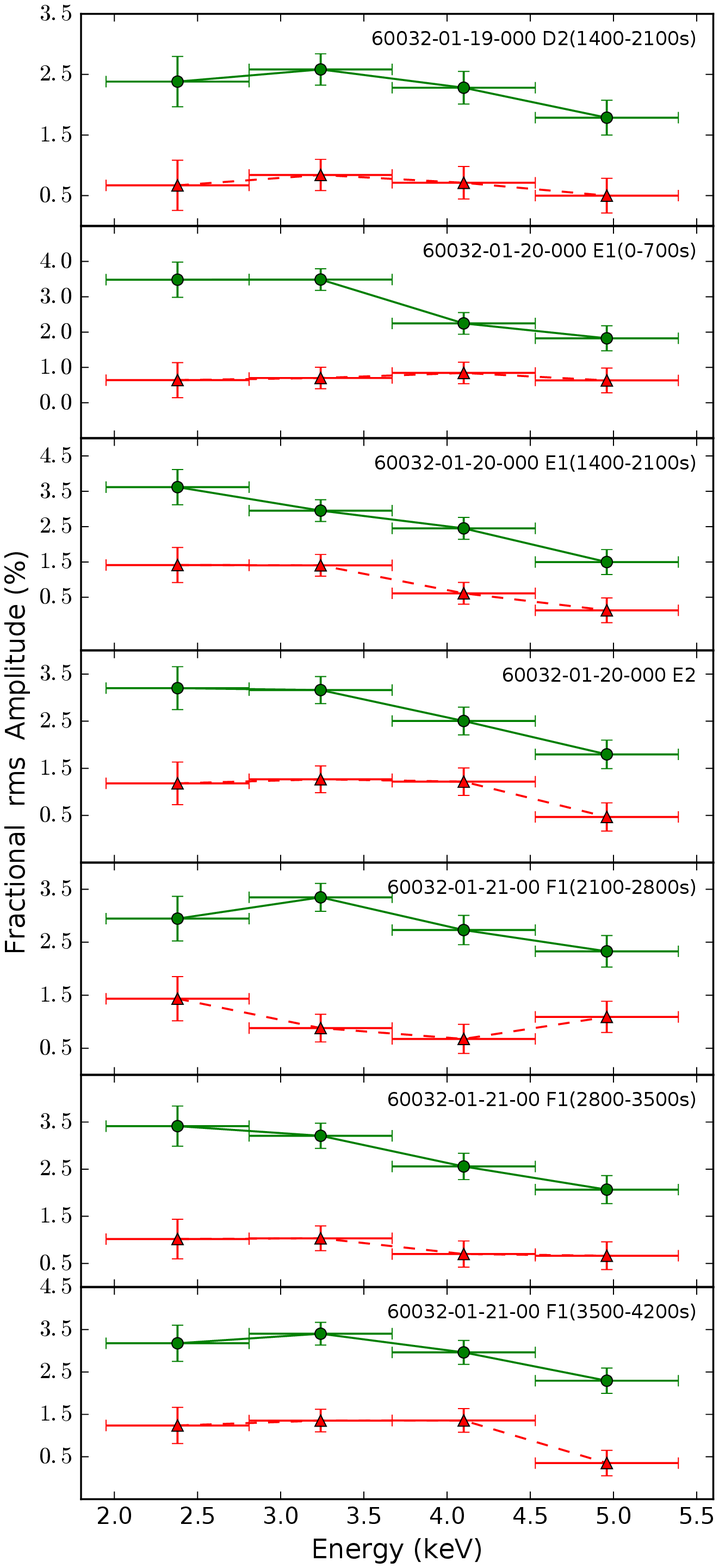}
\includegraphics[height=115mm,width=72mm]{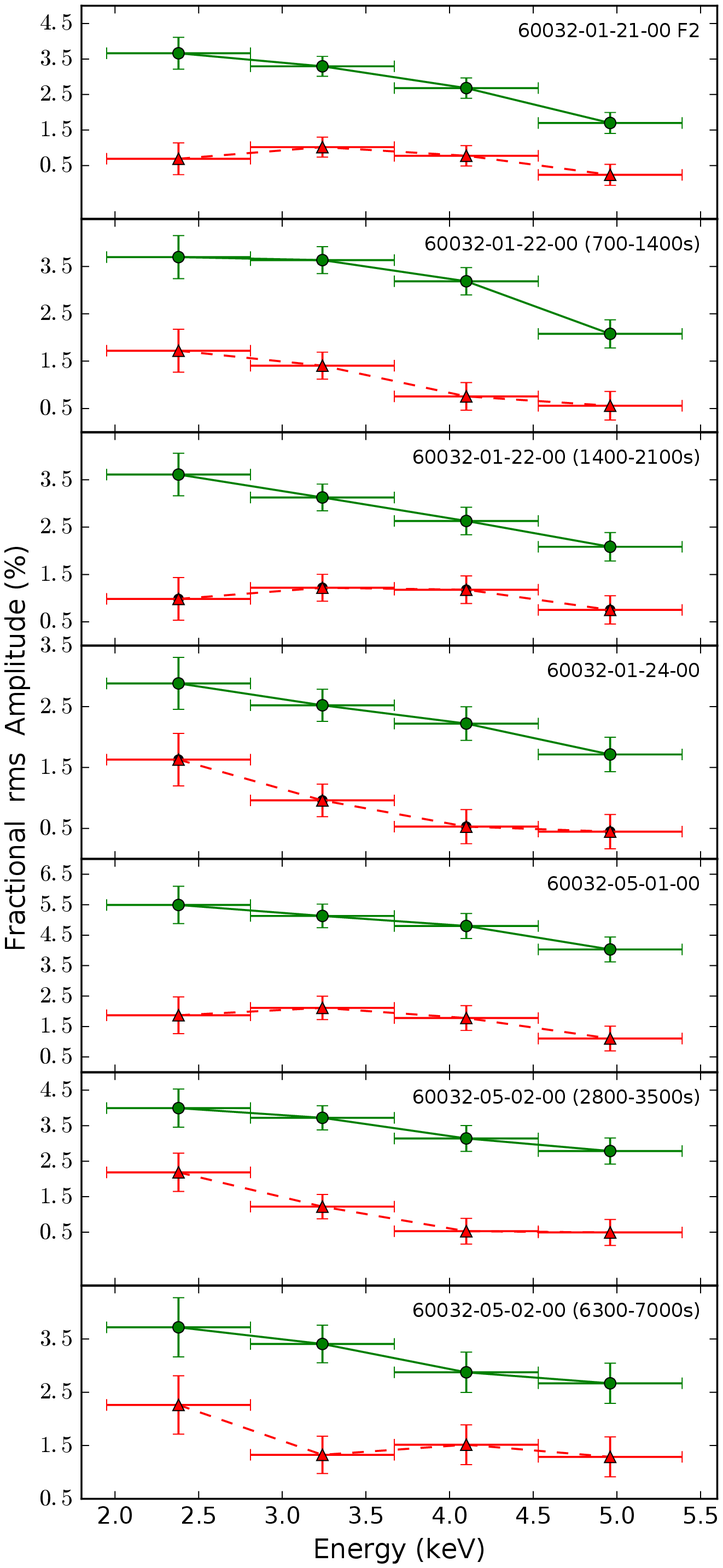}
\end{figure*}
\begin{figure*}
\centering
\includegraphics[height=115mm,width=72mm]{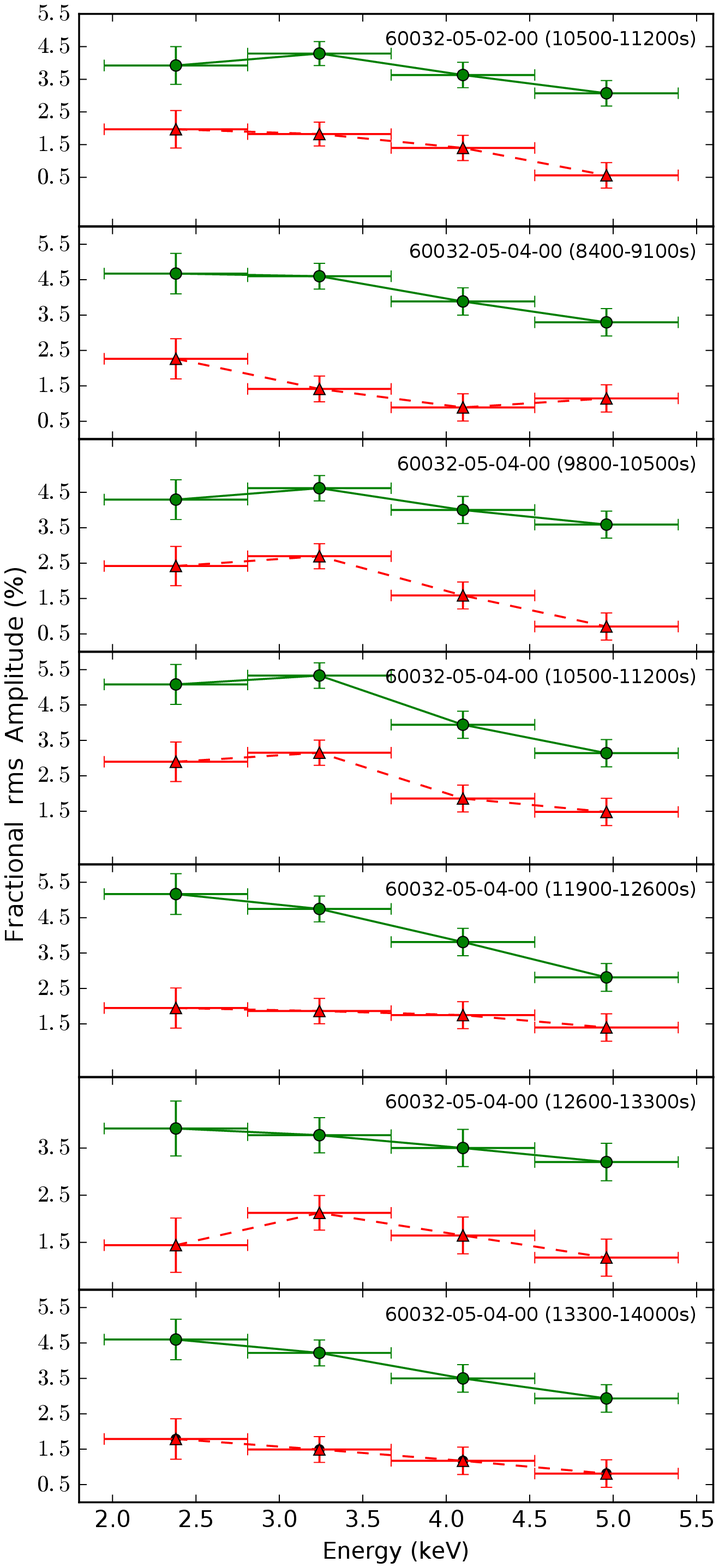}
\includegraphics[height=115mm,width=72mm]{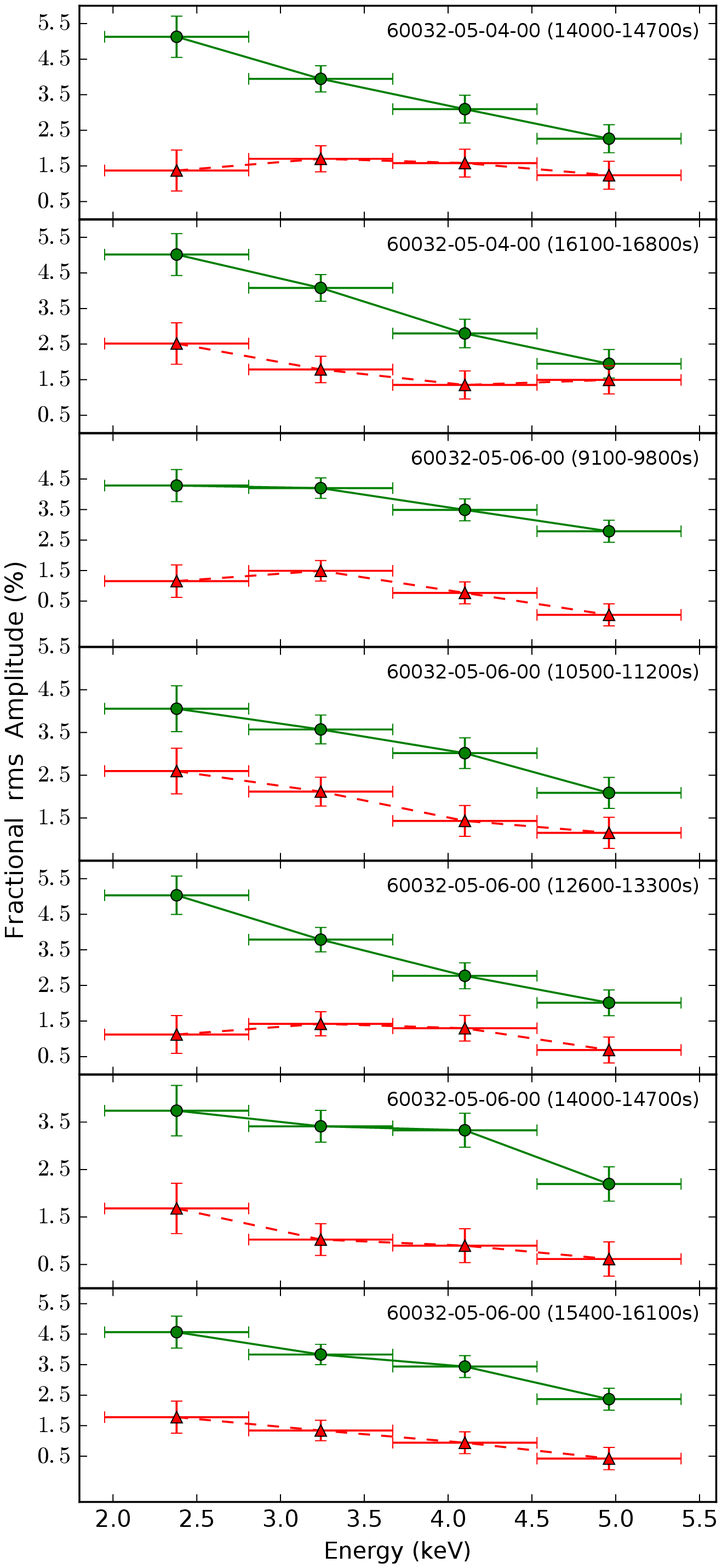}
\includegraphics[height=115mm,width=72mm]{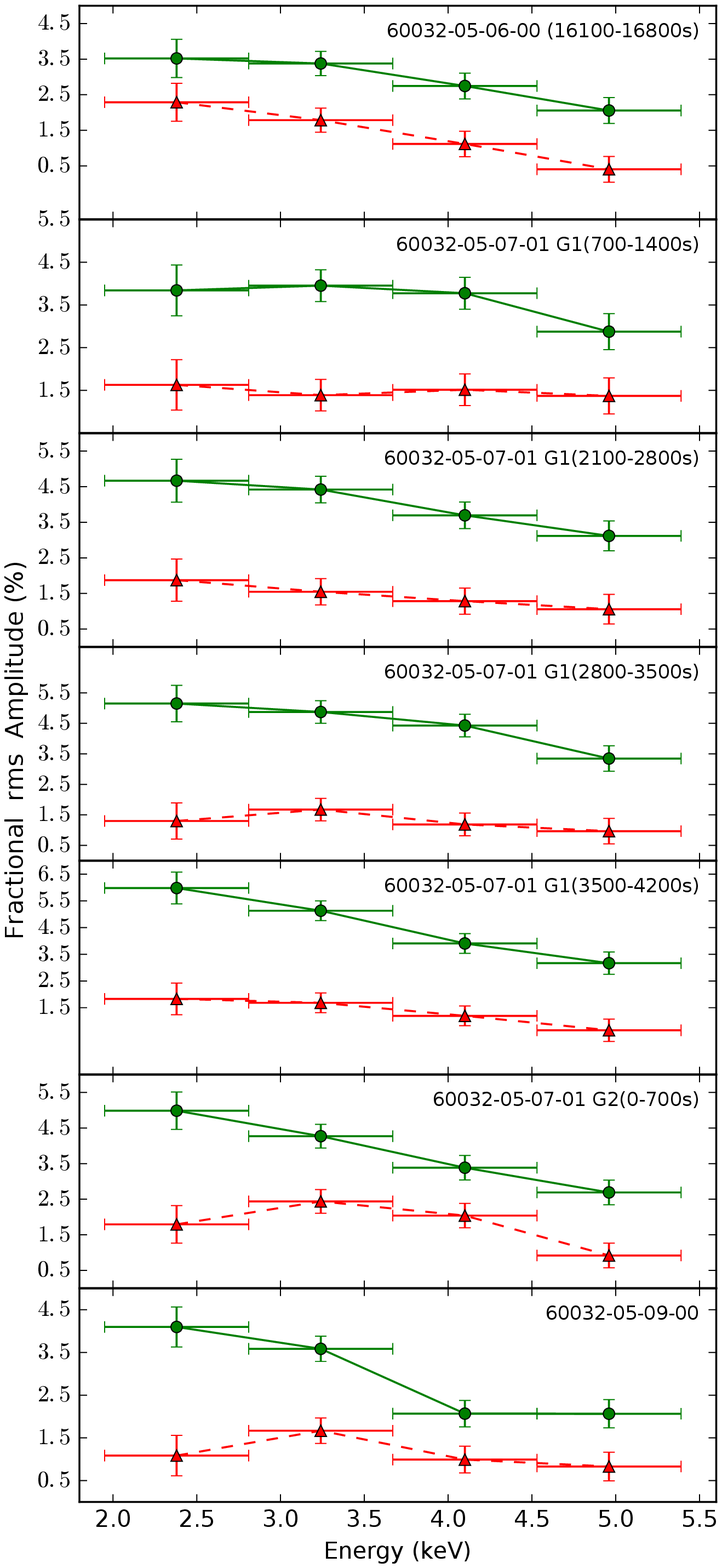}
\includegraphics[height=115mm,width=72mm]{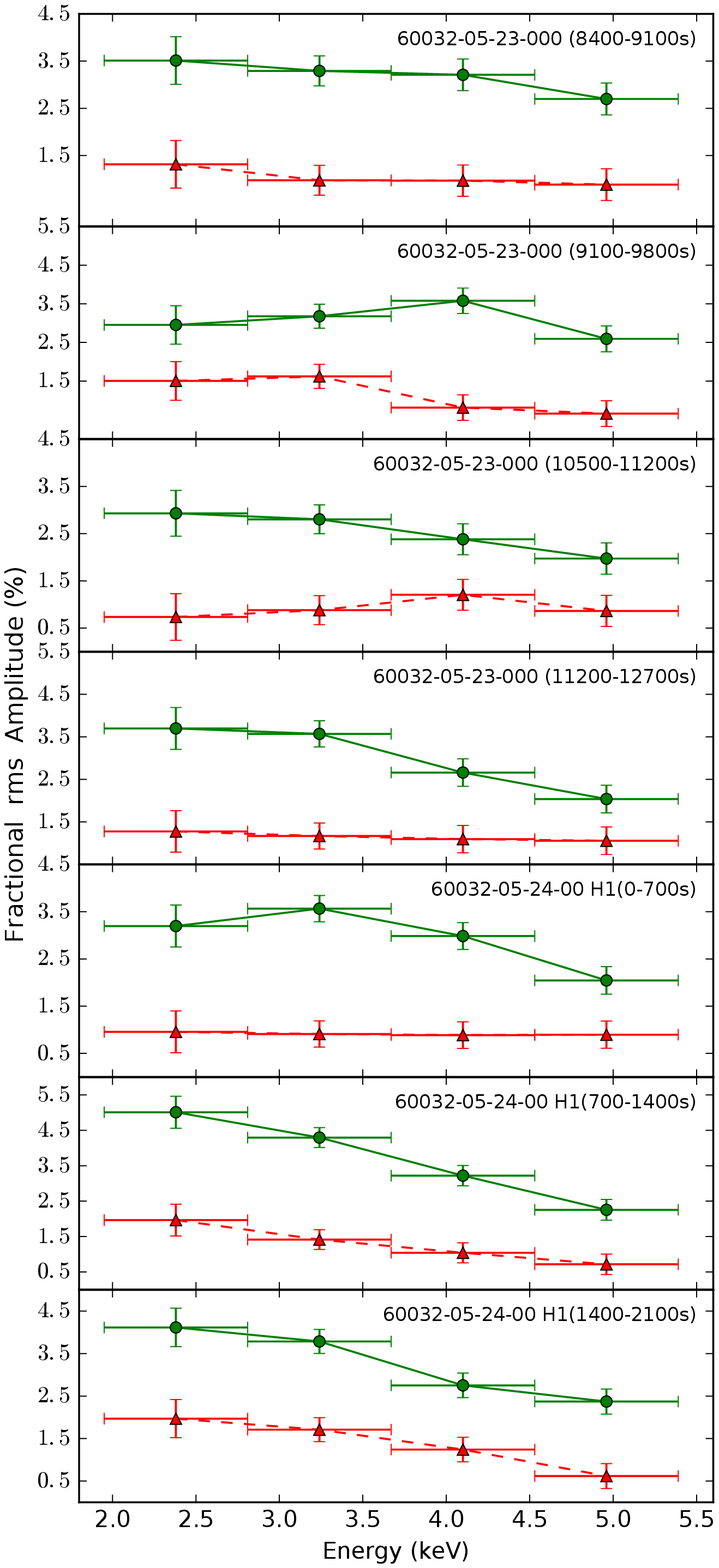}
\end{figure*}
\begin{figure*}
\centering
\includegraphics[height=115mm,width=72mm]{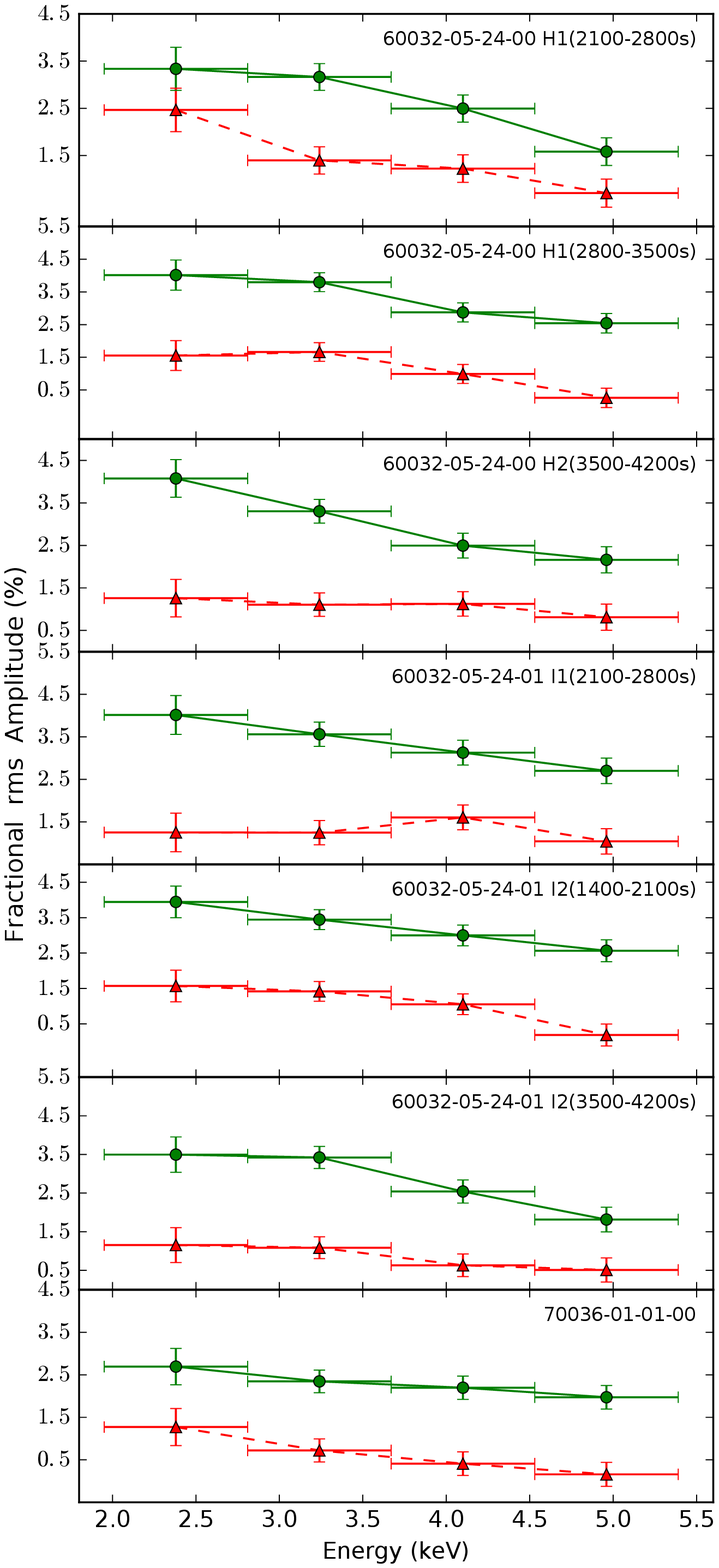}
\includegraphics[height=115mm,width=72mm]{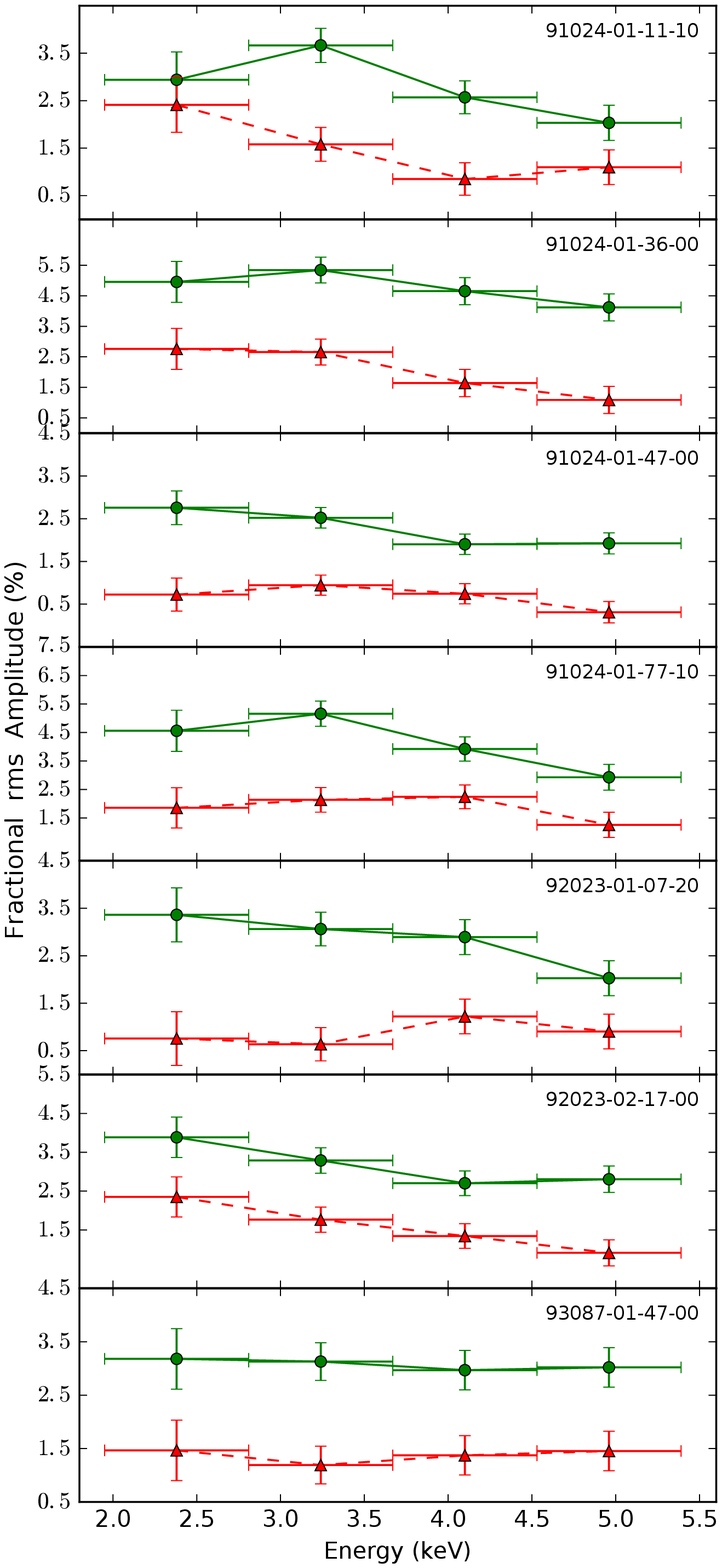}
\includegraphics[height=47mm,width=72mm]{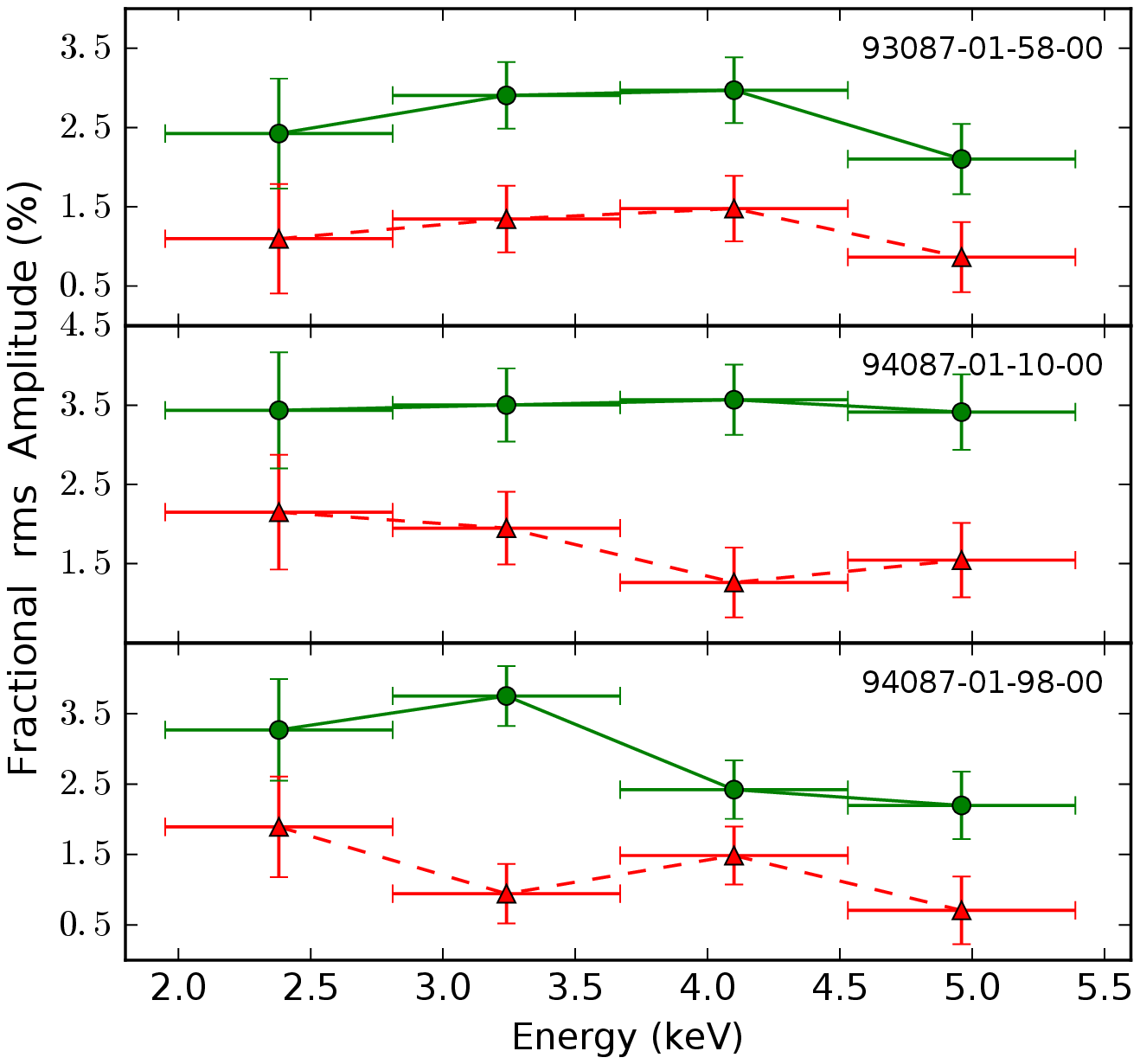}
\caption{Fractional rms amplitude of the fundamental (green circle) and the harmonic (red triangle) of the mHz QPO in 4U 1636--53 as a function of energy. We added an offset of 1.2\% to the rms of the fundamental to avoid overlap with that of the harmonic in each panel for comparison. The energy in the plot represents the central energy of each band, with the error bar indicating the energy range of the band.}
\label{res1}
\end{figure*}

\section{Discussion}  
We studied the fractional and absolute rms amplitude of the fundamental and the harmonic of the mHz QPO in 4U 1636--53 using RXTE observations. We found that the harmonic of the mHz QPOs appears when the source is in the transitional and the soft spectral state, with most of detections in the vertex in the color-color diagram. For the first time, we found that the ratio between the rms amplitude of the fundamental and that of the harmonic remains constant, $\sim$0.6. We also found that the rms spectrum of the harmonic follows the same trend as the one of the fundamental when the energy increases from $\sim$ 2 keV to $\sim$ 5 keV, although the mechanism behind the harmonic of the mHz QPOs remains unexplained.

\begin{figure*}
\centering
\includegraphics[width=0.53\textwidth]{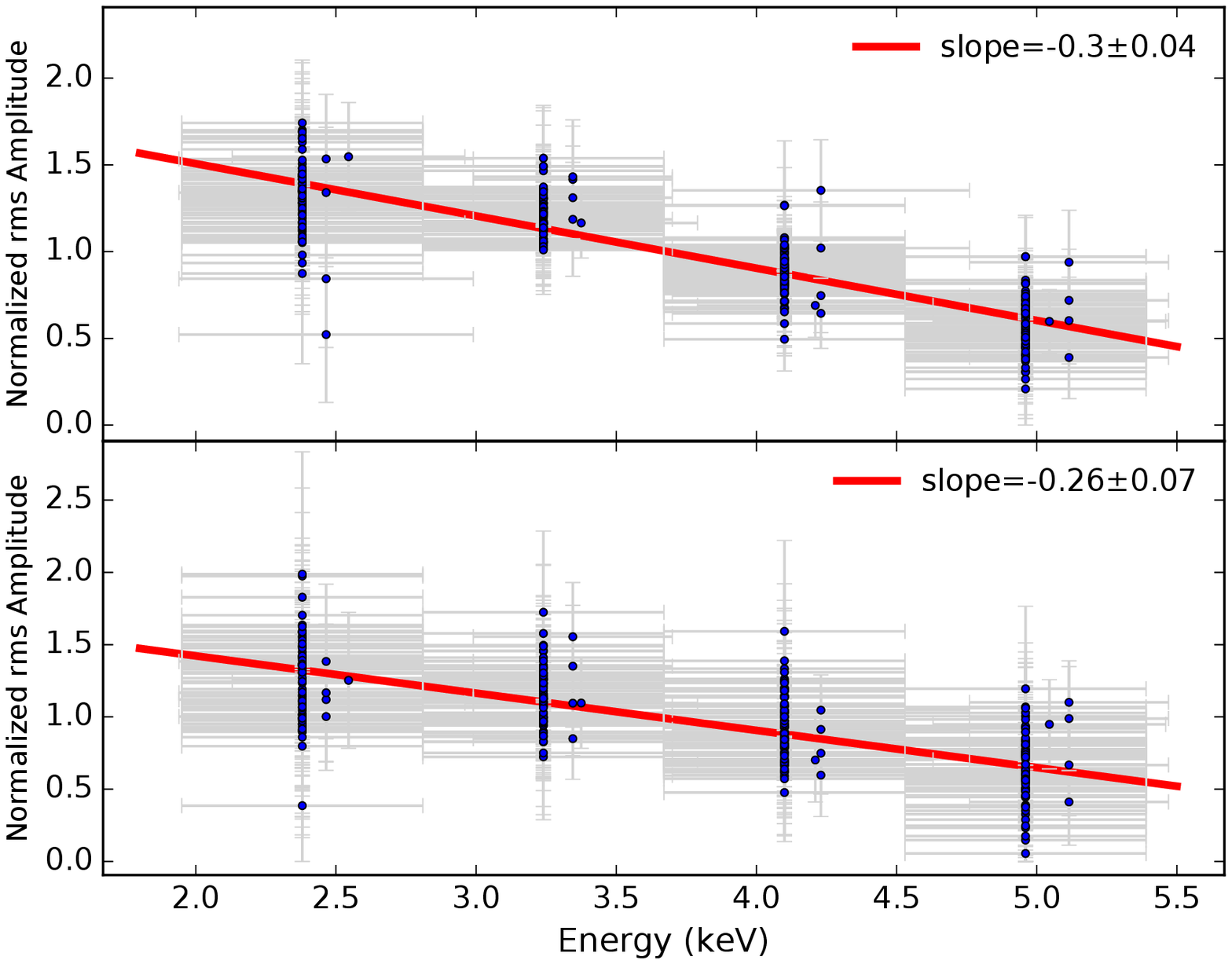}
\caption{Normalized rms amplitude of the fundamentals (upper panel) and the harmonics (bottom panel) of the mHz QPOs in 4U 1636--53 as a function of energy. We fitted each dataset with a linear relation and showed the fitted slope in the legend. The energy in the plot represents the central energy of each band, with the error bar indicating the energy range of the band.}
\label{sploe_nor}
\end{figure*}

\begin{figure*}
\centering
\includegraphics[width=0.45\textwidth]{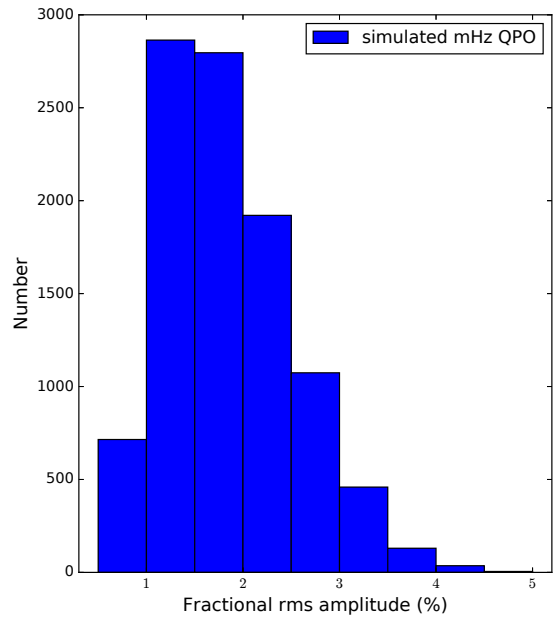}
\caption{Distribution of the fractional rms amplitude of $10^{4}$ simulated mHz QPOs. The rms distribution of the simulated QPOs is very close to the one of the mHz QPOs reported in \citet{Lyu2019} with RXTE data.}
\label{simulation}
\end{figure*}

\subsection{ Do all mHz QPOs have the harmonics ?}  

We detected the harmonic component of the mHz QPOs in 73 independent 700s intervals, much less than the number of intervals with mHz QPOs reported in \citet{Lyu2019}. We also calculated the rms amplitude of the possible harmonic in the intervals in which we did not detect it. The derived average rms amplitude and 1-$\sigma$ error of the harmonic in these intervals are 0.6\% and 0.2\%, respectively, indicating that the possible harmonic oscillations are on average 2-$\sigma$ different from zero. As a comparison, these numbers are 1.2\% and 0.2\% in the intervals where the harmonics are detected. The limited detection of the harmonics suggests that in 4U 1636--53 either most of the mHz QPOs do not have the harmonic component, or the harmonic is weak and cannot be detected in the intervals of 700 seconds exposure that we used in this work. 

Assuming that the waveform of the mHz oscillations is a sine curve, we further carried out a Monte Carlo simulation as follows to figure out whether all the mHz QPOs have the harmonics:

(1) We generated $10^{4}$ time series of 700s with a mHz QPO using the formula $A sin(2 \pi f t + \phi)+B$, where the term $A sin(2 \pi f t + \phi)$ and $B$ describe the mHz oscillations and the persistent rate, respectively. We determined the parameters $A$, $f$ and $B$ according to the results in \citet{Lyu2019}: the persistent rate $B$ comes from a normal distribution, with the mean value being a random number in the range covered by the rate of all the data in \citet{Lyu2019}, and the standard deviation being 5\% of the average. We calculated the parameter $A$ from the rms amplitude and the average rate with the formula $A=\sqrt{2}B*rms$, where $rms$ follows a Gaussian distribution with mean and standard deviation of, respectively, 1.25\% and 1\%, estimated from Figure 4 in \citet{Lyu2019}. The frequency $f$ was taken from a normal distribution, with the average and the standard deviation of the distribution at, respectively, 8.3 mHz and 1.1 mHz, the same as the ones in \citet{Lyu2019}. The phase, $\phi$, is taken as a random number in the range [0,2$\pi$] during the simulation. The rms amplitude distribution of the 10$^{4}$ simulated mHz QPOs (Figure \ref{simulation}) is similar to the one in the RXTE data reported in the work of \citet{Lyu2019}.

(2) Among the $10^{4}$ time series with the mHz QPOs, we then randomly selected $10^{3}$ time series to have harmonics. We added another sine function to describe the harmonic component in each time series. The frequency of the harmonic is assumed to be twice that of the fundamental frequency, and the phase is a random number between 0 and 2$\pi$. The amplitude of the harmonic is calculated from the amplitude of the fundamental and the ratio $rms_{h}/rms_{f} \sim 0.6$ derived in this work: $A_{h}=0.6*A_{f}$, where $A_h$ and $A_f$ are, respectively, the amplitude of the harmonics and the fundamentals.

(3) We then applied the Lomb-Scargle periodogram to search the harmonic component in these $10^{3}$ time series, and counted the number of detections at a 3-$\sigma$ confidence level or higher. 

(4) We repeated the steps (2)-(3) five times and calculated the average detection number in these five rounds, which would be a good indicator of the detection rate of the harmonics based on the data and the method used in this work.

For the five rounds, the detection fraction is between 47\% and 50\%, with the average detection rate being $\sim$ 48\%, suggesting that only half of the harmonics could be effectively detected at a confidence level of 3-$\sigma$ or higher. The loss of the other half of the harmonics in the segment with mHz QPOs could be due to, for instance, the limitation of the method, or the limited exposure of the 700s intervals. An interesting thing that we found in the simulations is that the detection of the harmonics not only depends on their rms amplitude, but also upon the phase difference between the harmonic and the fundamental. For a given phase difference between them, the peak power of the harmonic in the Lomb-Scargle periodogram increases as the rms amplitude of the harmonic increases. This trend, however, no longer holds when the phase difference is set to be random. In this case, the harmonics with high rms amplitude could have a low peak in the L-S periodogram, and thus be below the 3-$\sigma$ confidence level. 

With the results derived in this work, we found that not all the mHz QPOs in 4U 1636--53 have the harmonics:

(a) \citet{Lyu2019} systematically studied the properties of the mHz QPOs in 4U 1636--53 and found that the distribution of the fractional rms amplitude of the mHz QPOs ranges from $\sim$0.5\% to $\sim$4.5\%, with mHz QPOs in $\sim$ 100 intervals having an rms amplitude above 2\%. If we assume that the harmonics and the fundamentals are coupled and always appear together, then these 100 intervals should show a significant harmonic component since the calculated harmonic rms amplitudes would be larger than 1\%, according to the rms ratio of $\sim$ 0.6 derived in this work. However, as shown in Figure \ref{hist_Frac_rms}, here we detected significant harmonics only in $\sim$ 35 intervals, suggesting that the harmonics are not always present.

(b) With a detection rate of $\sim$ 48\%, we can deduce the total number of harmonics in 700s intervals in this work. We found that the number, 73$/$0.48=152, is $\sim$ 40\% of the total number of the intervals with mHz QPOs reported in \citet{Lyu2019}. This result is consistent with the finding that there is no clear harmonic component in one XMM-Newton observation in the dynamic power spectrum in \citet{Lyu2015}.

We could speculate about the possible causes for the appearance of the harmonics although there is no related description about the harmonics in the mHz QPO models at present. The harmonics could be present if the shape of the signal from the marginally stable burning itself is not always strictly sinusoidal. \citet{Vikhlinin1994} showed that the QPO and the harmonics could naturally appear in the framework of the shot noise model. To keep a stable release of energy, \citet{Vikhlinin1994} proposed that the appearance of strong shots should influence the amplitude or the probability of occurrence of the subsequent shots, hence leading to the peaks in the power density spectrum. Therefore, the harmonics of the mHz QPOs could appear when the shape of the signal from the marginally stable burning is not sinusoidal, for instance, in the form of random shots with rapid rise and exponential decay.

The scenario that the mHz QPO profiles are not always strictly sinusoidal is reasonable based on the current results. Simulations in the work of \citet{Heger2007} predicts that the profile of the oscillations in the light curve is not symmetric, instead, its decay part lasts twice as long as the rise. Observationally, the profiles of the mHz QPOs are not always the same: the peaks of the oscillation in the light curve at the frequency of the mHz QPOs in \citet{Revnivtsev2001} were asymmetric, with a steep rise and a shallower decline. Conversely, the average, phase-folded light curve of the mHz oscillation in GS 1826--238 by NICER in the recent work of \citet{Strohmayer2018} was quite symmetric. Therefore, in some cases the profile is sinusoidal, while in other cases the profile is not strictly sinusoidal. 

There are also other possible scenarios responsible for the harmonics. A harmonic component would mean that the oscillation pattern has some special symmetry. A pulsar with two polar caps will illuminate the observer twice per rotational cycle, thus producing a harmonic signal at twice rotational frequency. In the case of the marginally stable burning something similar may be happening. For instance, if the accreted matter in-falls onto the neutron star surface unevenly (influenced, e.g., by the magnetic field), it could likely reach two spherically symmetric places on the neutron star surface, triggering marginally stable burning at these two places and generating harmonics.

\subsection{ rms amplitude of the harmonics and the fundamentals}  
It is important to compare the rms spectra of the harmonics and the fundamentals to see whether the harmonics of the mHz QPOs have the same rms spectra as the fundamentals based on the detected harmonics in this work. There are already some reports in the literature where the rms vs energy relation of the harmonic is different from that of the fundamental. \citet{Cui1999} found that the rms of the harmonic component of the 67 mHz QPO in the microquasar GRS 1915+105, which increases as energy increases, is different from the one of the fundamental. \citet{Rao2010} investigated the low-frequency oscillations (LFQPOs) in XTE J1550--564, and found that the rms amplitude of the fundamental in type-C LFQPOs increases with energy, while that of the harmonic component first increases below a few keV, and then decreases as the energy further increases. \citet{Li2013MNRAS} reported a similar behavior of the fundamentals and the harmonics in this black hole system. The reason for the difference in the rms spectrum of these QPOs is still unclear, however, these inconsistencies cause confusions about the formation of those harmonics. In contrast, the derived rms amplitude of the harmonic and the fundamental in this work show a consistent trend as energy increases. There is no related description or prediction in the current mHz QPO models, but this finding favors a scenario in which the origin of the harmonic is consistent with that of the fundamental.

Besides, we noticed that the ratio of the rms amplitude between the harmonic and the fundamental remains constant regardless of the variation of the fundamental frequency and the soft count rate. This finding is different from the ratio in the low-frequency QPOs mentioned above. E.g., \citet{Rao2010} studied the low-frequency QPO in the black hole binary XTE J1550--564 and found that the rms amplitude of the harmonic peak divided by that of the fundamental decreases as the frequency of the fundamental increases. It indicates that the QPO signal in XTE J1550--564 becomes more sinusoidal as the fundamental frequency increases. Conversely, in this work we found that the rms ratio is constant, indicating that the rms amplitude of the fundamental and the harmonic are linearly correlated. It might be an evidence that the harmonic and the fundamental come from the same physical process. Furthermore, the constant ratio of 0.6 suggests that the oscillation profile is no longer a standard sine waveform due to the relative strong harmonic, and the shape of the signal from the marginally stable burning remains unaffected  even when there is a big change in the radiation luminosity or accretion rate.

\section{acknowledgements}

This research has made use of data obtained from the High Energy Astrophysics Science Archive Research Center (HEASARC), provided by NASA's Goddard Space Flight Center. This research made use of NASA's Astrophysics Data System. Lyu is supported by National Natural Science Foundation of China (grant No.11803025). D.A. acknowledges support from the Royal Society. G.B. acknowledges funding support from the National Natural Science Foundation of China (NSFC) under grant No. U1838116 and Y7CZ181002. GCM was partially supported by PIP 0102 (CONICET) and received financial support from PICT-2017-2865 (ANPCyT). F.Y.X. is supported by the Joint Research Funds in Astronomy (U2031114). X.J. Yang is supported by the National Natural Science Foundation of China (NSFC) under grant No. 11873041.

\acknowledgments

\bibliographystyle{aasjournal}
\bibliography{paper}

\end{document}